 \documentclass[manuscript]{aastex}
\shorttitle{Saturation of corotation resonances}
\shortauthors{Masset \& Ogilvie}
\begin{document}
\title{On the saturation of corotation resonances: a numerical study}
\author{F.\ S. Masset\altaffilmark{1} \and G.\ I. Ogilvie\altaffilmark{2}}
\affil{DSM/DAPNIA/SAp, Orme des Merisiers, CE-Saclay, 91191 Gif/Yvette Cedex, France}
\email{fmasset@cea.fr}
\affil{Institute of Astronomy, University of Cambridge, Madingley Road,
	Cambridge CB3 0HA, UK}
\email{gogilvie@ast.cam.ac.uk}
\altaffiltext{1}{Send offprint requests to F.\ S.\ Masset: fmasset@cea.fr}

\begin{abstract}
The torque exerted by an external potential on a two-dimensional
gaseous disk at non-co-orbital corotation resonances is studied by
means of numerical simulations.  The degree of saturation of these
resonances is important in determining whether an eccentric giant
planet embedded in a protoplanetary disk experiences an eccentricity
excitation or damping.  Previous analytic studies of the saturation
properties of these resonances suffered two important restrictions, as
they neglected: (i) the possible overlap between neighboring
first-order corotation resonances, and (ii) the fact that first-order
corotation resonances share their location with principal Lindblad
resonances.  We perform calculations restricted to one or two
resonances to investigate the properties of two neighboring
corotation resonances, as well as the properties of a corotation
resonance that overlaps a Lindblad resonance.  We find that these
properties hardly differ from the case of an isolated corotation
resonance, { which is found in a first step to agree with the
analytical theory}.  In particular, although the torque of two
neighboring corotation resonances may differ from the sum of the
torques of the corresponding resonances considered as isolated, it
never exceeds the sum of the fully unsaturated isolated corotation
resonances, and it saturates in a fashion similar to an isolated
resonance.  Similarly, the presence of an underlying Lindblad
resonance hardly affects the corotation torque, even if that resonance
implies a torque strong enough to significantly redistribute the
azimuthally averaged surface density profile, in which case the
corotation torque scales with the resulting vortensity gradient.  This
set of numerical experiments thus essentially validates previous
analytic studies. As a side result, we show that corotation libration
islands misrepresented by a mesh of too low resolution can lead to a
strongly overestimated corotation torque. This may be an explanation
why the eccentricity of embedded Jupiter-sized planets was never
observed to undergo an excitation in the numerical simulations
performed so far.
\end{abstract}


%

\section{Introduction}

Among the puzzles raised by the statistics of the recently discovered
extrasolar giant planets (hereafter EGPs) is the origin of their
eccentricities. The population of EGPs with periods larger than $\sim
10$~days displays an important scatter in eccentricity, which reaches
for some systems values as extreme as $0.9$, while most of the
eccentricities of longer-period planets are almost uniformly
distributed between~$0$ and~$0.6 - 0.7$.  Several explanations have
been envisaged to date in order to account for these eccentricities,
{ such as planet-planet interactions (Rasio \& Ford 1996, Ford et
al. 2001), external perturbers such as distant binary companions or
passing stars (Holman et al. 1997, Mazeh et al. 1997), or disk--planet
interactions (Goldreich \& Sari 2003, Ogilvie \& Lubow 2003 and
references therein). In this last case these interactions} endow the
giant protoplanets with non-vanishing eccentricities that remain after
the dispersal of the disk.  Here we examine some aspects of this tidal
excitation of the eccentricity of a giant protoplanet, by means of
restricted numerical simulations.  In order for these simulations to
describe as closely as possible the resonances excited by an eccentric
giant planet, it is useful to bear in mind their location and their
action on the eccentricity.  In Table~\ref{tab:reslist} we indicate
the ratio $\Omega/\Omega_p$ at the location of principal and
first-order resonances and the impact of these resonances on the
planet's eccentricity, where we denote by $\Omega$ the orbital angular
velocity of fluid elements in the disk, and by $\Omega_p$ the planet's
orbital frequency.  The qualification {\em fast} refers to first-order
terms of the potential that have the pattern frequency
$\Omega_p+\kappa_p/m$, where $\kappa_p$ is the planet's epicyclic
frequency and $m$ the azimuthal wavenumber, while the {\em slow}
denomination refers to first-order terms that have a pattern frequency
$\Omega_p-\kappa_p/m$.  In Table~\ref{tab:reslist} we neglect the
slight sub-Keplerianity of the disk due to the radial pressure
gradient, we also neglect the slight offset between the nominal
position of a Lindblad resonance and its effective position
(corresponding to the turning point in the WKB dispersion relation of
density waves in the disk), and we exploit the Keplerian degeneracy
$\Omega_p=\kappa_p$ and $\Omega=\kappa$.

\placetable{tab:reslist}

In the case of an embedded giant protoplanet that clears a gap around
its orbit, none of the co-orbital resonances (the principal corotation
resonances, the fast first-order outer Lindblad resonances and the
slow first-order inner Lindblad resonances) contributes to the
exchange of energy and angular momentum with the planet, and the
eccentricity budget of the planet is therefore determined by the
extreme Lindblad resonances (the fast ILR and the slow OLR), which
excite the eccentricity, and by the first-order corotation resonances,
which damp it. It has long been known that the damping provided by the
CRs slightly exceeds the excitation due to the LRs. Goldreich \&
Tremaine (1980), using the large-$m$ Bessel-function approximation for
the torques and treating~$m$ as a continuous variable (an
approximation which might not be accurate for the gap-opening giant
planets which involve low $m$-values), have shown that the corotation
resonances dominate over the Lindblad resonances by a $5$\% amount.
The corotation resonances can however easily saturate, unlike the
Lindblad resonances.  If they reach a $5$\% saturation, the net effect
on the eccentricity is an excitation. Ogilvie \& Lubow (2003) and
Goldreich \& Sari (2003) have studied the saturation properties of an
isolated corotation resonance. The saturation of a corotation
resonance results from the competition between libration (which tends
to flatten the vortensity profile across the libration islands of the
resonance) and viscous diffusion (which tends to establish the
large-scale gradient of vortensity over the width of the libration
islands).  The saturation of a resonance therefore requires a
sufficiently small viscosity, for a given resonance width. For a given
mass and eccentricity of the planet, there exists a critical viscosity
below which the planet undergoes an eccentricity excitation, at least
according to the balance between the resonant interactions. Ogilvie \&
Lubow (2003) show that for a planet-to-primary mass ratio $q=0.001$,
and a disk with aspect ratio $H/r=0.05$, a starting eccentricity of
order $e=0.01$ is sufficient to trigger an eccentricity growth.  Their
analysis however neglected the fact that non-co-orbital corotation
resonances are not isolated. Successive corotation resonances can
indeed overlap, and these resonances share their location with the
principal Lindblad resonances, as can be seen in
Table~\ref{tab:reslist}.

This work aims at studying the properties of a non-isolated corotation
resonance by means of numerical simulations. Our analysis is {
threefold: we first check that numerical simulations of an isolated
corotation resonance provide results in agreement with the analytical
theory, then} we investigate the properties of a pair of corotation
resonances of similar strength with wave-numbers that differ by~$1$,
from the well separated case to the strongly overlapping case, and,
finally, we investigate the behavior of a first-order corotation
resonance that lies at the location of a principal Lindblad
resonance. Disentangling the torque from each resonance in that last
case is made possible by the fact that the azimuthal wavenumber of the
Lindblad resonance and that of the corotation resonance differ by~$1$,
as can be seen in Table~\ref{tab:reslist}.

\section{Numerical issues and setup}
\subsection{Numerical scheme}
The scheme is one commonly used in the context of the tidal
interaction of a protoplanetary disk and an embedded planet (Nelson
\& al. 2000, d'Angelo \& al. 2003, Bate \& al. 2003).  It is an upwind
scheme on a staggered polar mesh, in which the slopes are computed
using a second-order estimate (van Leer 1977, Stone \& Norman 1992).
This kind of scheme is known to exhibit a spurious post-shock
oscillatory behavior that is regularized by the use of a second-order
artificial viscosity. In the present work however, in which we hardly
ever had to deal with shocks (which appear only in the case of a
strong overlapping Lindblad resonance), this artificial viscosity
plays no role.  We work in a frame rotating with angular speed
$\Omega_f$, which corotates either with one of the resonances or with
the mid-radius of our annular domain. We use the angular momentum
conservative scheme described by Kley (1998) for a rotating frame,
rather than considering the Coriolis force as a source term.  A
significant speed-up is obtained using the fast azimuthal advection
method described by Masset (2000).  In a standard manner, we consider
a locally isothermal 2D gaseous disk with sound speed $c_s(r)$ ($r$
being the distance to the mesh center), which we characterize by its
aspect ratio $H/r=c_s(r)/[r\Omega(r)]$, assumed to be uniform over the
computational domain.  The disk is torqued by a potential
$\Phi(r,\phi,t)$ that corresponds either to one or two rotating
components:
\begin{equation}
\label{eq:pot}
\Phi(r,\phi,t)=T(t/\tau)\sum_{i=1}^{1\mbox{~\footnotesize or~}2}
\psi_i(r)\cos[m_i(\phi-\Omega_it)],
\end{equation}
where $\phi$ denotes the azimuthal angle, $t$ the time, $\psi_i(r)$
the radial profile of the $i^{th}$ component of the potential, $m_i$
its azimuthal wavenumber, $\Omega_i$ its pattern speed, and where
\begin{eqnarray*}
T(x)&=&\left[\sin\left(\frac\pi 2x\right)\right]^2\mbox{~if~}x<1\\
	&=&1\mbox{~~~~otherwise}
\end{eqnarray*}
is a temporal tapering that turns on the potential on the
time-scale~$\tau$.
\subsection{Annular domain width}
In order for the torque arising from each component of
equation~(\ref{eq:pot}) to correspond only to a specific resonance,
some care must be taken about the radial width over which the disk is
torqued.  One can either adopt a profile $\psi_i(r)$ that is
non-vanishing on a radial range smaller than that of the computational
domain, or one can adopt a constant $\psi_i(r)$ profile and a
sufficiently narrow computational domain, so as to remain everywhere
sufficiently far from the other resonances associated to that
potential component (e.g.  from the Lindblad resonances in the case we
want to consider a corotation resonance). We have adopted this second
solution, and we work on radially narrow meshes.  Namely, we consider
low-$m$ resonances ($m=2$~or~$3$). For the less favorable case $m=3$,
the LR to CR radius ratios are $r_{OLR}/r_{CR}=1.21$ and
$r_{ILR}/r_{CR}=0.76$.  We want the computational domain to avoid the
LR locations, at least by typically two or three driving lengths
(Meyer-Vernet \& Sicardy 1987):
\[\lambda_D\simeq\left(\frac{H^2r}{3m}\right)^{1/3}.\]
This implies that we have a sufficiently thin disk (the ``standard''
choice $H/r=0.05$, for which the driving length is $\lambda_D\sim
0.06r$, would give marginally too large Lindblad resonance widths,
since the corotation resonance --~which is at least as large as
$H$~-- would slightly overlap the perturbation inside of the
OLR). Instead, we use a much thinner disk ($H/r=0.01$) for which the
wave driving length at the Lindblad resonance is approximately three
times shorter.  With this choice, we can safely consider two distinct
corotation resonances that are well separated from the Lindblad
resonances.

\subsection{Boundary conditions}
Far from an isolated corotation resonance, but still inside of the
band forbidden for the propagation of acoustic perturbations around
corotation, the perturbation corresponds to an evanescent wave, with
a purely imaginary radial wave-number given by:
\[k_{r_i}^2=\frac{1}{c_s^2}[m_i^2(\Omega-\Omega_i)^2-\Omega^2]-\frac{m_i^2}{r^2}\]
obtained from the dispersion relation of a density wave in a Keplerian
disk. In the case where we consider only corotation resonances, we
evaluate this wave-number for each resonance $i$ at the mesh
boundaries, and we apply the following procedure:
\begin{itemize}
\item 
We evaluate the azimuthal Fourier transform of the hydrodynamic
variables (surface density, radial and azimuthal velocities) with
wavenumber $m_i$, in the second ring at the inner boundary and in the
penultimate ring at the outer boundary:
\begin{eqnarray*}
A_{m_i}&=&\frac 2{N_\phi}\sum_{k=0}^{N_\phi-1}\xi_k\cos\left(\frac{2\pi}{N_\phi}m_ik\right)\\
B_{m_i}&=&\frac 2{N_\phi}\sum_{k=0}^{N_\phi-1}\xi_k\sin\left(\frac{2\pi}{N_\phi}m_ik\right),
\end{eqnarray*}
where $(\xi_k)_{0\leq k < N_\phi}$ is any of the aforementioned
quantities on the corresponding ring, and where $N_\phi$ is the number
of zones of the mesh in azimuth.
\item 
We then overwrite the content of the first ring (at the inner
boundary) and of the last ring (at the outer boundary) with:
\begin{eqnarray*}
\xi_k=&\xi^0+\sum_i\left[A_{m_i}\cos\left(\frac{2\pi}{N_\phi}m_ik\right)\right.+\\
&\left.B_{m_i}\sin\left(\frac{2\pi}{N_\phi}m_ik\right)\right]\exp(-|k_{r_i}|\Delta r),
\end{eqnarray*}
where $\Delta r$ is the radial resolution of the mesh, and where
$\xi^0$ stands for the unperturbed hydrodynamic variable.
\end{itemize}
This simple prescription ensures a correct description at the mesh
boundaries of the evanescent waves arising for all the corotation
resonances.

In the case where one considers both a corotation resonance and a
Lindblad resonance, a similar procedure does not produce satisfactory
results, and the computational domain is rapidly invaded by spurious
propagating waves. We then turn to a procedure proposed by Kley,
Nelson \& Artymowicz for a comparison test problem\footnote{\tt
http://www.astro.su.se/$\sim$pawel/planets/test.hydro.html}, in which
the perturbation is artificially damped on a short range of radius
near either boundary.  Although this prescription mangles the response
of the disk to a corotation resonance near its boundaries, it is very
effective at avoiding the reflection of the wave launched at the
Lindblad resonance.

\subsection{Centrifugal balance}
The mechanism we want to investigate relies on a minute motion inside
libration islands, which involves very small radial velocities. The
idealized picture of these libration islands can easily be mangled by
an inaccurate initial centrifugal balance, which gives rise to
axisymmetric waves with an associated radial velocity which can be
larger than the one corresponding to libration. It is therefore
critical to impose a strict centrifugal balance at $t=0$, which takes
into account the discretization of the problem on the mesh. Since the
mesh is staggered there is an infinite number of sequences
$(v_\phi^j)_{0\leq j<N_r}$ (where $N_r$ is the radial number of zones
of the mesh) that lead to a strict centrifugal balance. We choose the
one that minimizes the quantity $R$ defined by:
\[R=\sum_{j=1}^{N_r-2}|v_\phi^{j+1}+v_\phi^{j-1}-2v_\phi^j|\]

\subsection{Torque expression}

The total torque exerted by the potential on the disk is evaluated as:

\[\Gamma=\sum_{i=0}^{N_r-1}\sum_{j=0}^{N_\phi-1}
	-\frac{\Phi_{ij+1}-\Phi_{ij-1}}{2\Delta\phi}S_{ij}\Sigma_{ij},
\]
where $S_{ij}$ is the surface area of zone $(i,j)$, $\Sigma_{ij}$ and
$\Phi_{ij}$ respectively the surface density and gravitational
potential pertaining to the center of this zone, and
$\Delta\phi=2\pi/N_\phi$ the resolution in azimuth.

We shall also use the torque exerted on the disk by the specific
potential component with azimuthal wavenumber $m$. Its expression is
given by :

\begin{eqnarray*}
\Gamma_m&=&\sum_{i=0}^{N_r-1}\sum_{j=0}^{N_\phi-1}
S_{ij}\Sigma_{ij}\frac{\sin(m\Delta\phi)}{\Delta\phi}\times\\
&&[A_m^i\sin(mj\Delta\phi)-B_m^i\cos(mj\Delta\phi)]
\end{eqnarray*}
where

\begin{eqnarray*}
A_m^i&=&\frac{2}{N_\phi}\sum_{j=0}^{N_\phi-1}\Phi_{ij}\cos(mj\Delta\phi)\\
B_m^i&=&\frac{2}{N_\phi}\sum_{j=0}^{N_\phi-1}\Phi_{ij}\sin(mj\Delta\phi).
\end{eqnarray*}

The value of $\Gamma_m$ coincides exactly with the value of $\Gamma$
whenever only one potential component is present, regardless of the
resolution.

\subsection{Units}
As in a large number of publications on the disk--planet tidal
interaction problem, the mass unit is the mass of the central object,
the distance unit is arbitrary (in what follows it either corresponds
to the corotation radius of an isolated resonance, or the mean of the
corotation radii of two resonances), and the time unit is
$\Omega^{-1}$, where $\Omega$ is the Keplerian angular speed at $r=1$.
The surface density is initially uniform and equal to one.

\section{The isolated corotation resonance}
In a first step we have checked what resolution was needed for the
code to properly capture the torque properties (unsaturated torque
value and saturation properties) in the case of an isolated
corotation resonance. For this purpose, we have made a number of test
calculations with different resolutions and viscosities.  Our first
setup consists of an isolated $m=3$ resonance at frequency $\Omega=1$.
The inner mesh boundary is located at $R_{\rm min}=0.93$, and the
outer mesh boundary is located at $R_{\rm max}=1.07$.  The potential
is turned on on a time-scale $\tau=150$~$\Omega^{-1}$.  The number of
zones in $\phi$ is set to $90$, which means that each libration island
spans $30$~zones azimuthally.  The width of the libration islands is
given, if one neglects pressure effects, by
\begin{equation}
\label{eq:width}
W=(32)^{1/2}\delta_\psi=\frac 8\Omega\left(\frac\psi 3\right)^{1/2},
\end{equation}
where 

\[\delta_\psi=\left(\frac{2\psi}{3\Omega^2}\right)^{1/2}\]
is the characteristic length-scale for libration in a Keplerian disk
(Ogilvie \& Lubow 2003).  The forcing potential amplitude is
$\psi=10^{-5}$, so the width of the libration islands, neglecting
pressure effects, is $W\simeq 0.014$.  This typically corresponds to a
Jupiter-sized planet orbiting a solar-type star on an eccentric orbit
with $e \sim 4\cdot 10^{-3}$.  We have tried to vary the number of
zones radially, from $N_r=30$ to $N_r=240$.  This corresponds to a
radial sampling of the libration islands that varies from $3$ to
$25$~zones.  In this last case, the number of zones over which a
libration island is described is approximately the same in azimuth and
radius. The kinematic viscosity is either set to $0$, in which case
diffusion is only of numerical origin, or to the constant value
$10^{-6}$, which corresponds to a small ($\sim 7$\%) saturation of the
resonance. This can be evaluated by using the saturation parameter 
introduced by Ogilvie \& Lubow (2003), the expression of which, in a 
Keplerian disk, is

\begin{equation}
  \label{eq:sat}
  p=\left(\frac{2\psi}{3\Omega^2}\right)
  \left(\frac{3m\Omega}{2\nu r}\right)^{2/3},
\end{equation}
which leads in our case to the value $p=0.182$. The $7$\% saturation
is then deduced using equation~(66) of Ogilvie \& Lubow (2003), which
holds for slightly saturated resonances.

\begin{figure} 
\plotone{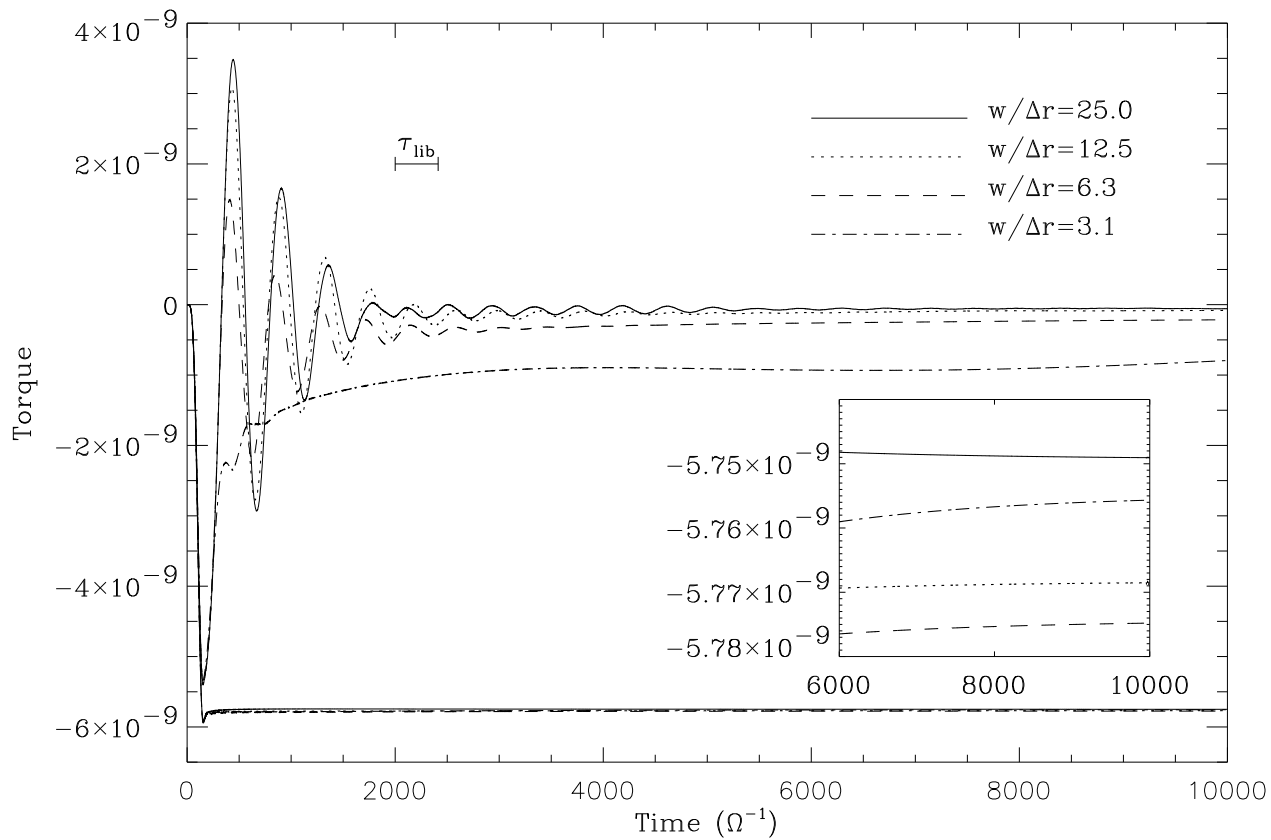}
\caption{\label{fig:one}%
Corotation torque exerted by the external potential on the disk as a
function of time, for an inviscid disk (upper curves) or for a viscous
disk (lower curves), for different radial resolutions.  The curves
corresponding to the viscous case differ very little from each other,
as can be seen in the close-up.}
\end{figure}

\placefigure{fig:one}

The results are presented in Figure~\ref{fig:one}.  One can notice an
oscillatory behavior of the torque value in the inviscid case, with a
period of the order of the libration time

\[\tau_{\rm lib} \sim \frac{8r}{3\Omega\delta_\psi m}.\]

The torque value at large viscosity is in correct agreement with the
estimate that can be deduced from the Goldreich \& Tremaine (1980)
formula that leads to:
\[\Gamma_{\rm unsaturated} = -5.92\cdot 10^{-9}\]
The expected torque value is therefore $7$\% smaller in absolute
value, i.e. $\Gamma=-5.51\cdot 10^{-9}$. The error in the torque
given by our calculation is therefore typically $5$\%.  One can also
notice that the unsaturated torque value is very well described even
at low radial resolution (bottom curves).  This had already been
noticed by Masset (2002) in the context of the horseshoe region
torque.  On the contrary, the curves of different resolutions
noticeably differ in the inviscid case. The expected behavior is that
the torque tends toward zero (saturates) as time goes to infinity. The
three highest resolutions seem to be in reasonable agreement with this
expectation, within a small offset.  The case of lowest resolution,
however, does have a significant residual torque at large $t$, and
does not exhibit any oscillatory behavior.  It is likely that at
these low resolutions the numerical diffusion itself acts to
unsaturate the torque. One can also note that the oscillatory
behavior of the torque, in the inviscid case, increases as the
resolution increases.

\begin{figure}
\plotone{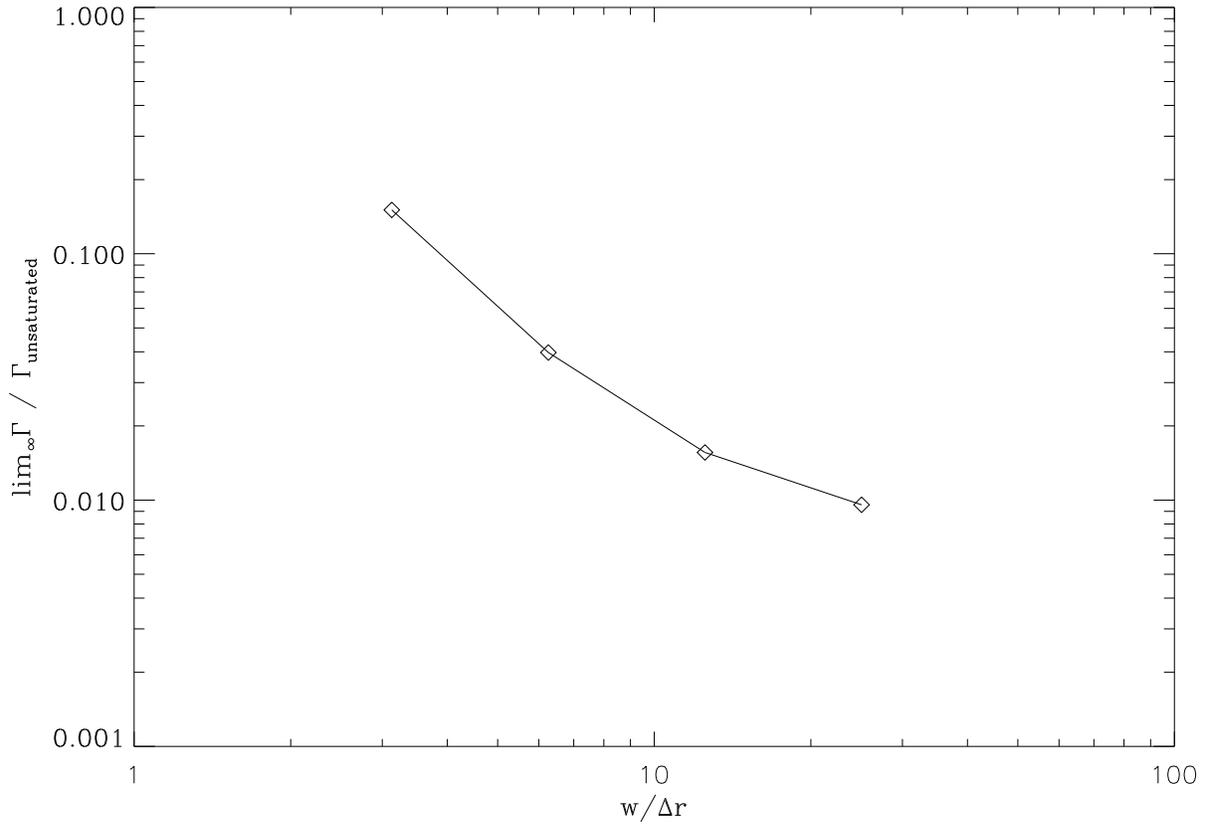}
\caption{\label{fig:two}%
Dimensionless residual corotation torque as a function of resolution,
for an inviscid disk.}
\end{figure}

\placefigure{fig:two}

In Figure~\ref{fig:two} we plot the dimensionless residual torque
value at large time as a function of the resolution, expressed as the
number of zones spanned by the libration islands. We see that a
resolution $w/\Delta r$ of $12$ to $25$ is sufficient to ensure a
residual unsaturation of about $1$\% after approximately 1000~orbits.

\begin{figure}
\plottwo{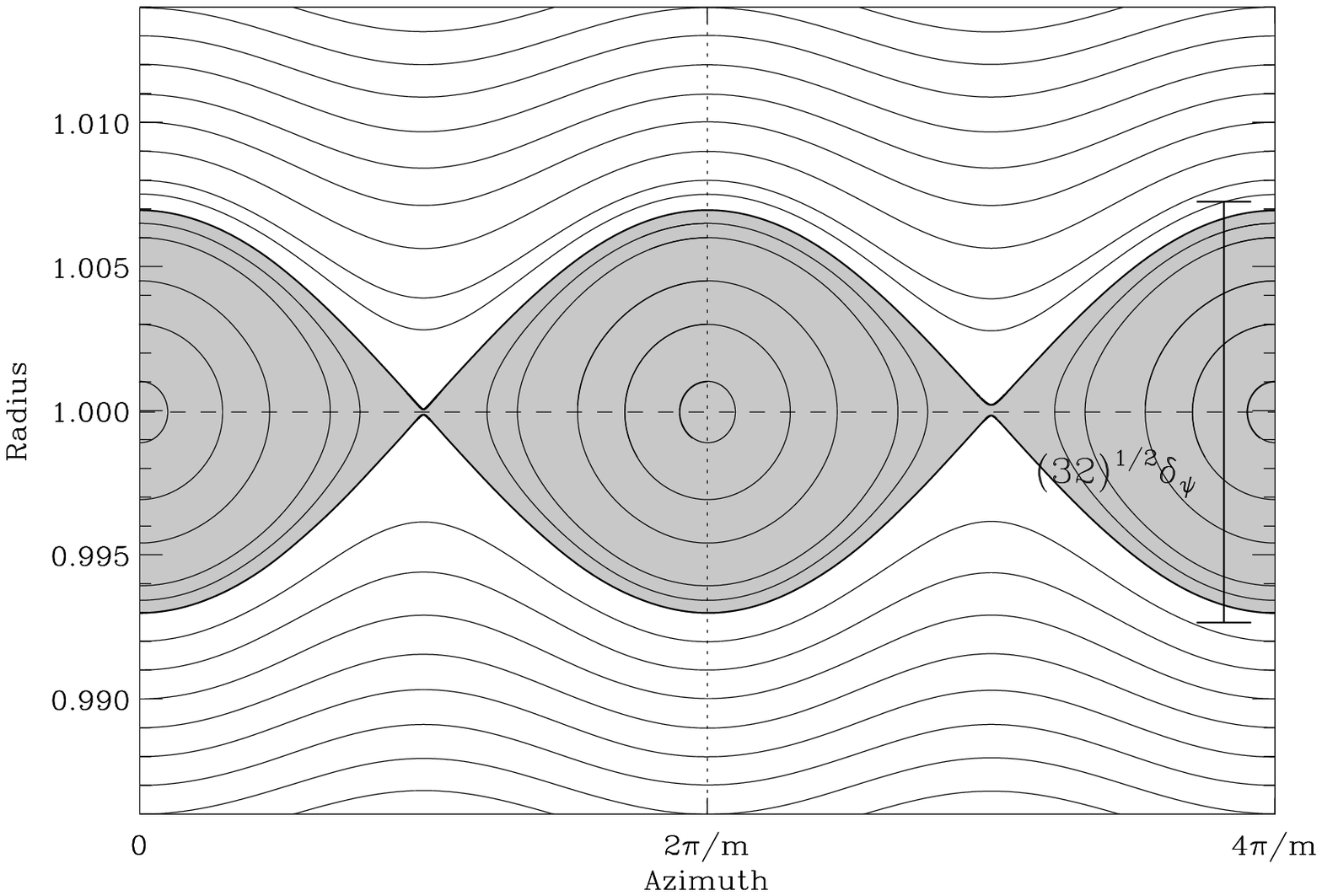}{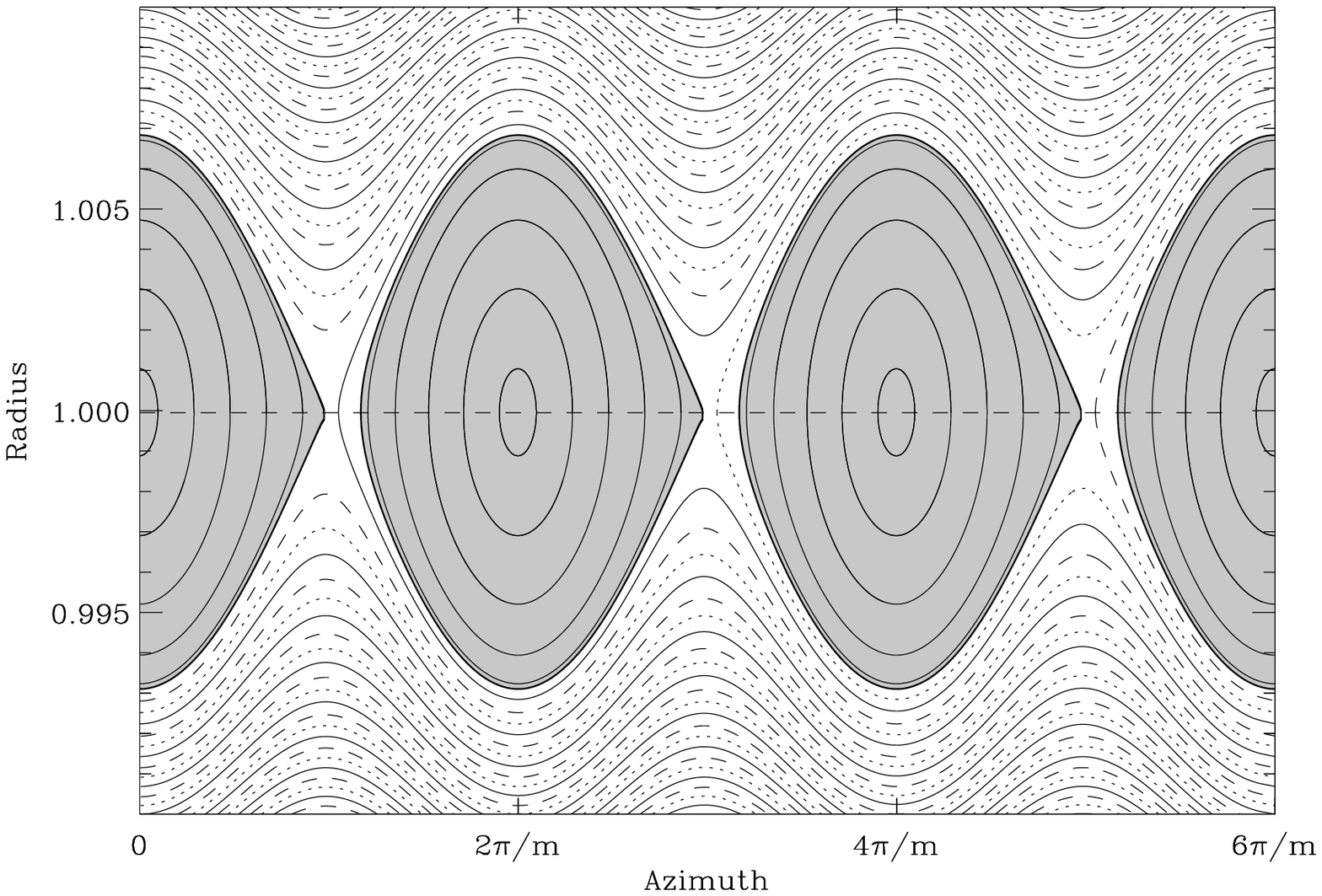}
\caption{\label{fig:three}Topology of the streamlines in the inviscid 
case (left panel) and for the viscous case (right panel). In the right
panel, the solid, dotted and dashed lines represent three distinct
unique streamlines that show how a fluid element originating in the
outer disk eventually reaches the inner disk after passing through the
corotation radius between the libration islands. This flow topology
is very similar to the one of the horseshoe region in a viscous disk
(Masset 2001, 2002).}
\end{figure}

\placefigure{fig:three}

In Figure~\ref{fig:three} we plot the streamlines (from the
calculation with highest radial resolution) in the inviscid case and
in the viscous case. As we consider a thin disk, for which the
pressure length-scale is of the order of the libration scale, the
radial width of the libration islands is satisfactorily accounted for
by the ballistic expression of equation~(\ref{eq:width}).

We have performed additional calculations for the isolated corotation
resonance case in order to investigate whether the viscosity
prescription ($\nu = \mbox{constant}$ or $\nu\propto\Sigma^{-2/3}$, as
used in the analysis of Ogilvie \& Lubow 2003), and the boundary
condition prescription, could significantly alter the value and
saturation properties of the corotation torque.

\begin{figure}
\plotone{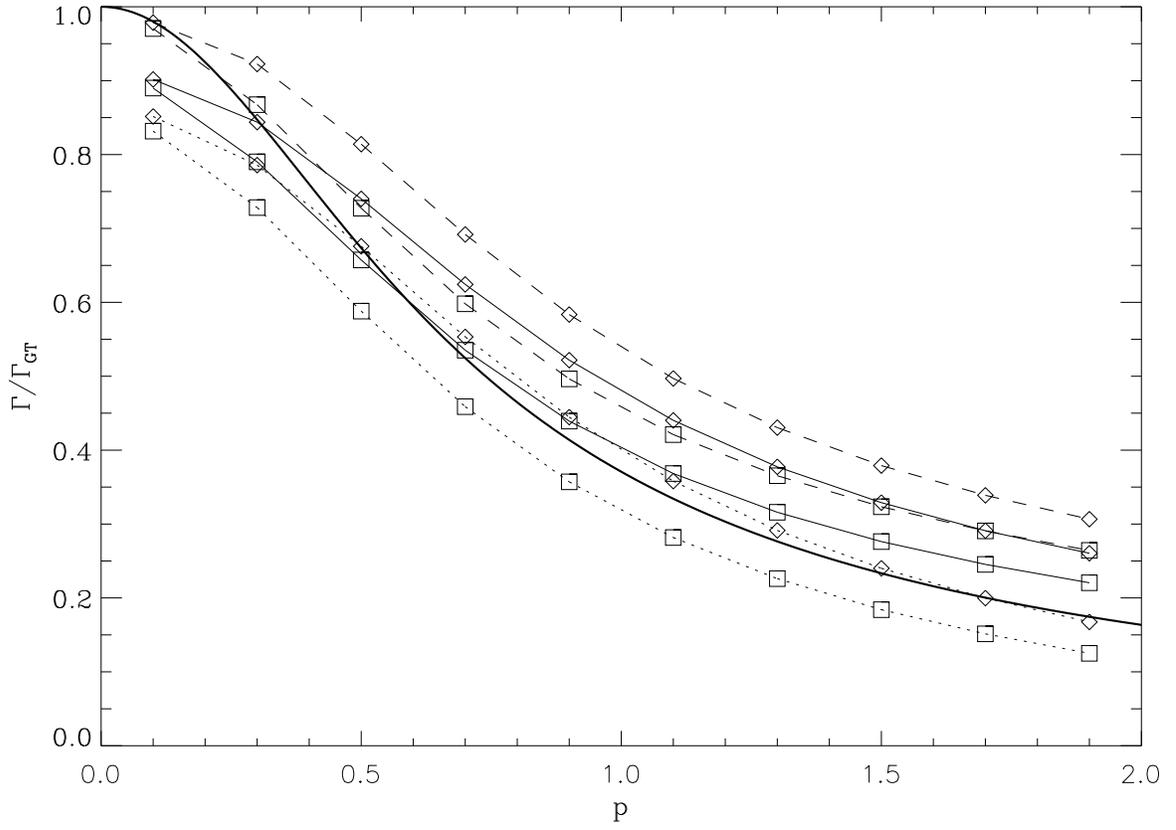}
\caption{\label{fig:four}%
Dimensionless corotation torque value of an isolated $m=3$ resonance, as a function
of the saturation parameter.
The thick solid line shows the theoretical curve for
the case $\nu\propto\Sigma^{-2/3}$ (Ogilvie \& Lubow 2003).}
\end{figure}

\placefigure{fig:four}

The results are shown in Figure~\ref{fig:four}.  The curves with
diamonds show the results with a constant viscosity, while the curves
with squares show the results with the viscosity prescription
$\nu\propto\Sigma^{-2/3}$.  All these calculations have $W/\Delta r =
21.5$ or~$22$. The solid lines show calculations with an inner
boundary at $R_{\rm min}=0.88$ and an outer boundary at $R_{\rm
max}=1.12$, with damping boundary conditions on the $7$\% inner and
outer parts of the mesh.  The dotted lines show calculations with the
same boundary conditions but with an inner boundary at $R_{\rm
min}=0.83$ and an outer boundary at $R_{\rm max}=1.17$.  Finally the
dashed lines show calculations with boundary conditions adapted to the
evanescent waves.  The dashed and dotted lines strictly have all same
parameters, and differ only by their boundary conditions.  One can
make the following comments:
\begin{itemize}
\item   
There is a systematic offset between a uniform viscosity prescription
and the $\nu\propto\Sigma^{-2/3}$ prescription, which amounts to a bit
less than 10\%, the latter prescription leading to a more saturated
resonance.
\item
The calculation that one would expect to best fit the analytical
results of Ogilvie \& Lubow (2003) is the one with adapted boundary
conditions and a $\nu\propto\Sigma^{-2/3}$ prescription. This
corresponds to the dashed line with squares. Although it
satisfactorily agrees with the theoretical expectations for $p \leq
0.3$, it leads to a resonance that is less saturated than expected for
larger values of $p$. For $p=2.0$, the measured dimensionless torque
is offset of $0.1$ from its analytical estimate. Such an offset is
unlikely to be due to numerical diffusion, as can be seen in
Fig.~\ref{fig:two}.
\item 
Damping boundary conditions have a small but measurable impact on the
fully unsaturated torque value.  This might be linked to the fact that
with the radially extended mesh that we consider, the boundaries lie
within $\sim 2$ driving lengths from the Lindblad resonances. In any
case, whenever we shall study a situation with two resonances (either
two corotation resonances or one corotation resonance and one
Lindblad resonance), we will compare our results to the isolated
corotation resonance case by performing separately an isolated resonance
test calculation, with exactly the same boundary location and
prescription.
\end{itemize}

\section{Saturation of a pair of corotation resonances}

We have performed a first set of calculations in which we consider the
$m=3$ corotation resonance presented at length in the previous section
($\psi=10^{-5}$, $\Omega=1$), and a second resonance, with potential
amplitude $\psi'=1.3\cdot 10^{-5}$, with azimuthal wavenumber $m'=2$,
and with frequency $\Omega'$ such that the corotation radius is
located at $r=1+s\cdot10^{-4}$, where $s$ takes the following values:
$25,50,100,200,300,500$.  With the largest separation ($s=500$), the
resonances are well apart and their distance in radius amounts to
slightly more than $3$ times the sum of their half-widths, while for
the smallest separation, the distance between the corotation radii
amounts to only 16\% of the sum of their half-widths, so the
resonances strongly overlap. For each two-resonance calculation, two
other separate calculations with exactly the same parameters are
performed, in which one only of the resonances is present.  We have
performed all these calculations with three different values for the
viscosity (chosen here as uniform): $\nu = 10^{-6}$, $\nu=10^{-7}$ and
$\nu = 0$.  The first choice corresponds to a low value of the $p$
parameter, and to resonances that are therefore almost unsaturated,
while the second choice corresponds to values of $p$ that are close to
unity ($p=0.844$ for the $m=3$ resonance, and $p=O(1)$ for the $m'=2$
resonance as well, with an exact value that depends on $s$). The mesh
parameters were $R_{\rm min}=0.87$, $R_{\rm max}=1.18$, $N_r=270$,
$N_\phi=90$.

\subsection{High viscosity}

\begin{figure}
\plotone{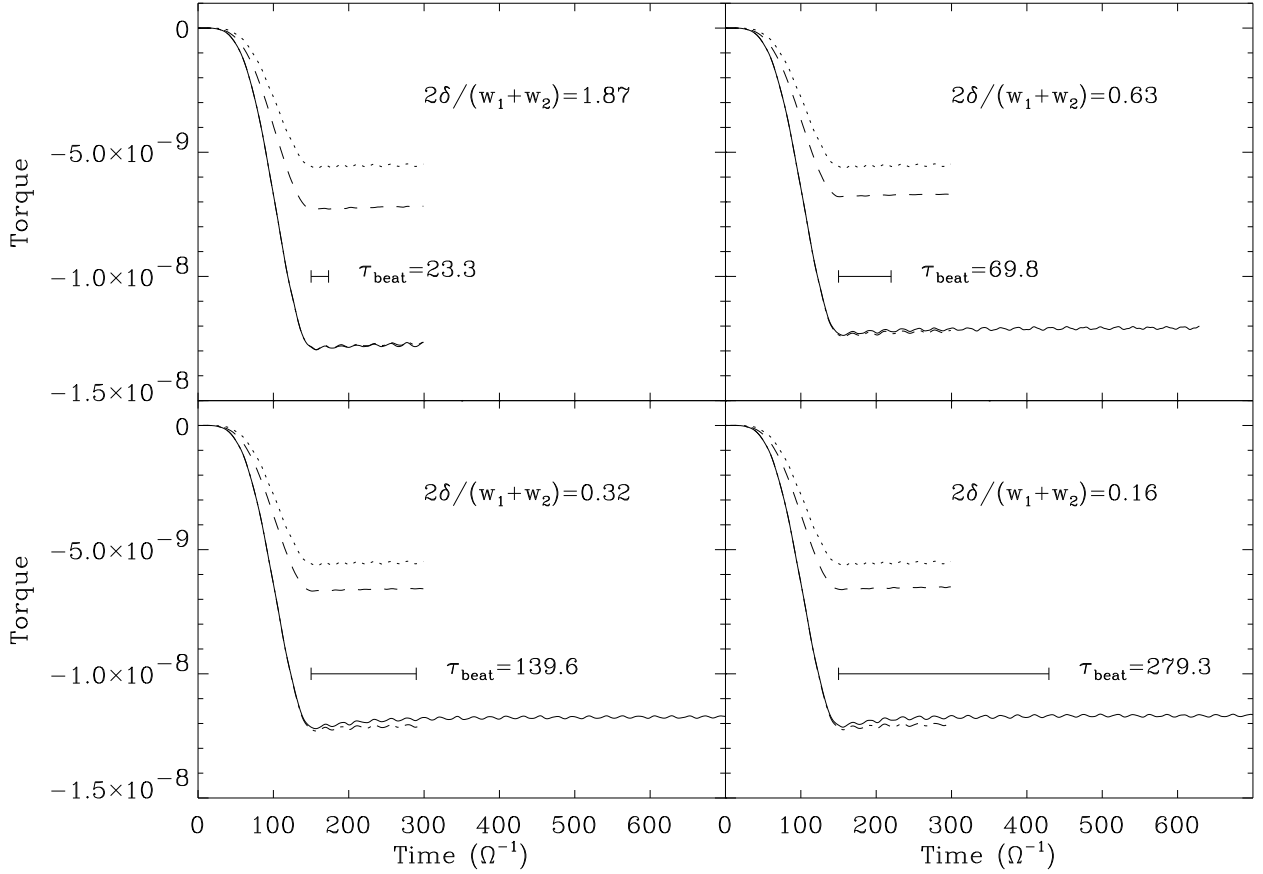}
\caption{\label{fig:five}%
Corotation torque of a pair of corotation resonances (solid line) in
the high-viscosity case, as a function of time.  The dotted line shows
the corotation torque of the isolated $m=3$ resonance, which is the
same on all plots since this resonance is fixed.  The dashed line
shows the corotation torque of the isolated $m'=2$ resonance, and the
dot-dashed line the sum of these one-resonance corotation torques. On
the first two plots, this curve and the solid curve are
indistinguishable by eye, while a slight offset can be noticed in the
strongly overlapping cases (bottom plots). It can be noticed that the
two-resonance corotation torque adopts an almost constant value over
a time that is larger than $\tau_{\rm beat}$. One can note a fast
oscillation of the torque value with a small amplitude. These
oscillations are found on the two-resonance runs as well as on the
isolated $m'=2$ runs, so they are probably due to the fact that the
$m'=2$ potential component and the mesh do not corotate.}
\end{figure}

\placefigure{fig:five}

In Figure~\ref{fig:five} we show results for four values of the
separation ($s=300,100,50,25$) of the highly viscous case.  On these
plots the variable $\delta$ represents the distance between the
corotation radii of the two resonances, $w_1$ is the $m=3$ resonance
width, and $w_2$ is the $m'=2$ resonance width, estimated with the
ballistic approximation of equation~(\ref{eq:width}). The resonances
should overlap when the parameter $\overline\delta=2\delta/(w_1+w_2)$
becomes smaller than unity, i.e. when the distance between the
corotation radii is smaller than the sum of the resonances'
half-widths.

The variable $\tau_{\rm beat}$ features in Figure~\ref{fig:five}.  It
represents the amount of time that separates identical configurations
of potential components:
\begin{equation}
\tau_{\rm beat}=2\pi\left|\frac{m^{-1}-m'^{-1}}{\Omega-\Omega'}\right|
\end{equation}

When the two resonances overlap, a passive scalar advected by the flow
displays a highly time-dependent behavior, and remarkably, in this
highly viscous case, this has no impact on the torque value.  The
two-resonance torque value is almost insensitive to the resonance
overlap, and exhibits a value that is nearly constant in time,
whatever the value of $\overline\delta$.  { A tentative explanation
for this property could be the following: Ogilvie \& Lubow (2003) have
emphasized that the nonlinear effect on the corotation torque of an
isolated resonance can be understood as the feedback of the
perturbation on the vortensity gradient of the disk, and this is
likely generalizable to a set of corotation resonances. For a
viscosity large enough that the viscous timescale across the libration
islands be shorter than any other timescale (libration timescale or
beating timescale), the vortensity gradient across the resonances is
hardly modified with respect to its unperturbed value, hence nonlinear
effects should be negligible and the torque value should be the sum of
the linear estimates for each resonance.}

\subsection{Moderate viscosity}
\begin{figure}
\plotone{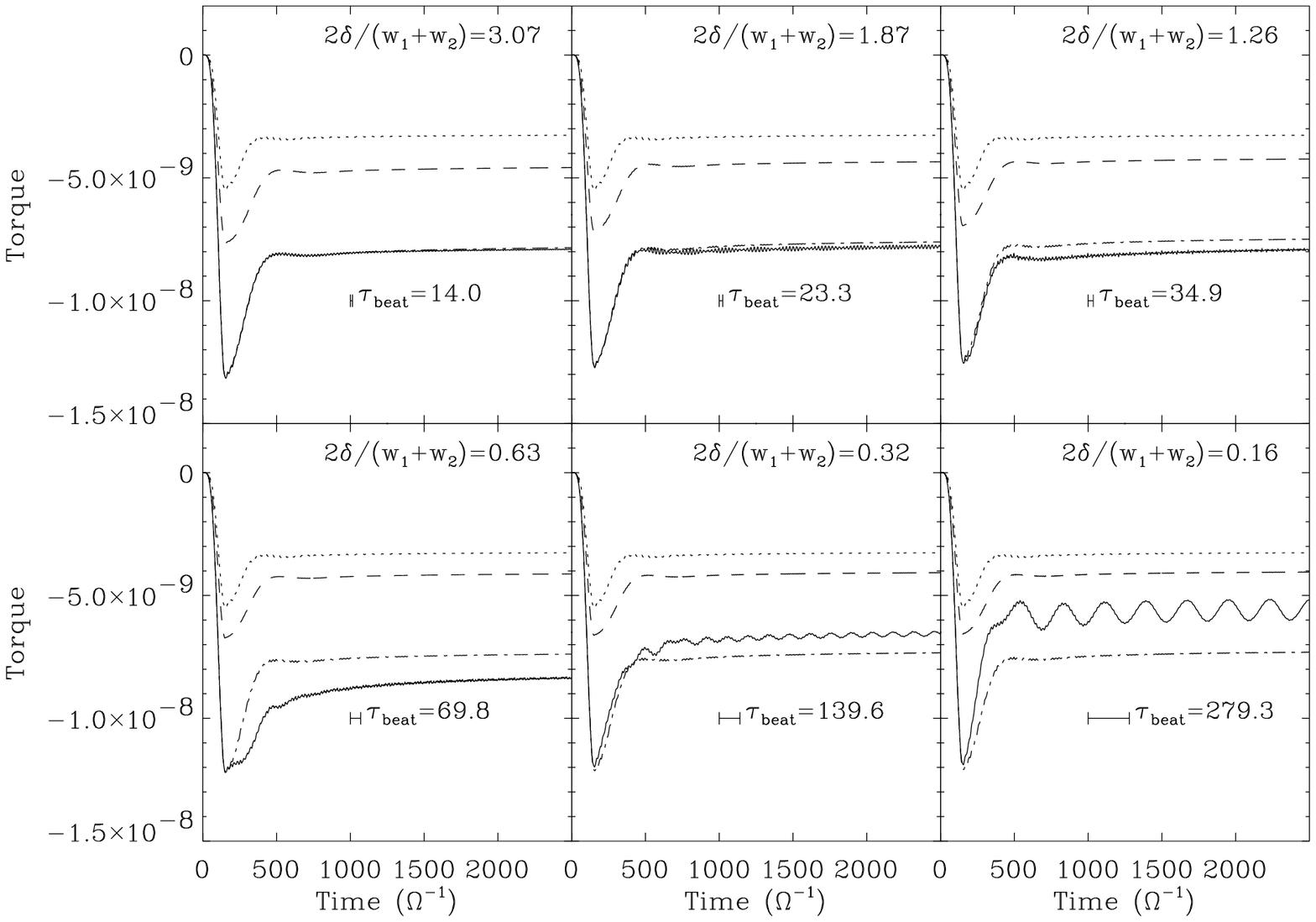}
\caption{\label{fig:six}%
Corotation torque of a pair of corotation resonances in the
moderate-viscosity case, as a function of time. The curve styles are the
same as in the previous figure.}
\end{figure}

In Figure~\ref{fig:six} we show results for six values of the
separation ($s=500,300,200,100,50,25$) of the moderately viscous case.
On these curves one can see that the two-resonance torque is nearly
equal to the sum of the one-resonance torques when the resonances are
separated ($\overline\delta > 1$), although there exists a small but
noticeable difference between these two values that increases as
$\overline\delta$ gets close to unity.  When both resonances overlap
(bottom panels), the time-averaged two-resonance torque value can be
either larger (in absolute value, for $\overline\delta=0.63$) or
smaller (for $\overline\delta=0.32$ or $\overline\delta=0.16$) than
the sum of the one-resonance torque values.  One can also notice a
torque modulation with period $\tau_{\rm beat}$.  The findings of the
highly viscous case therefore do not hold, even for the time-averaged
values. Note that $\tau_{\rm beat}$ is only the shortest period over
which one can expect the torque modulation to occur, while the longest
is $2\pi/[\mbox{GCD}(m,m')|\Omega-\Omega'|]$, where $\mbox{GCD}(m,m')$
is the greatest common divisor of $m$ and $m'$, which is $1$ in the
case where $|m'-m|=1$.

\placefigure{fig:six}

\subsection{Inviscid case}
\begin{figure}
\plotone{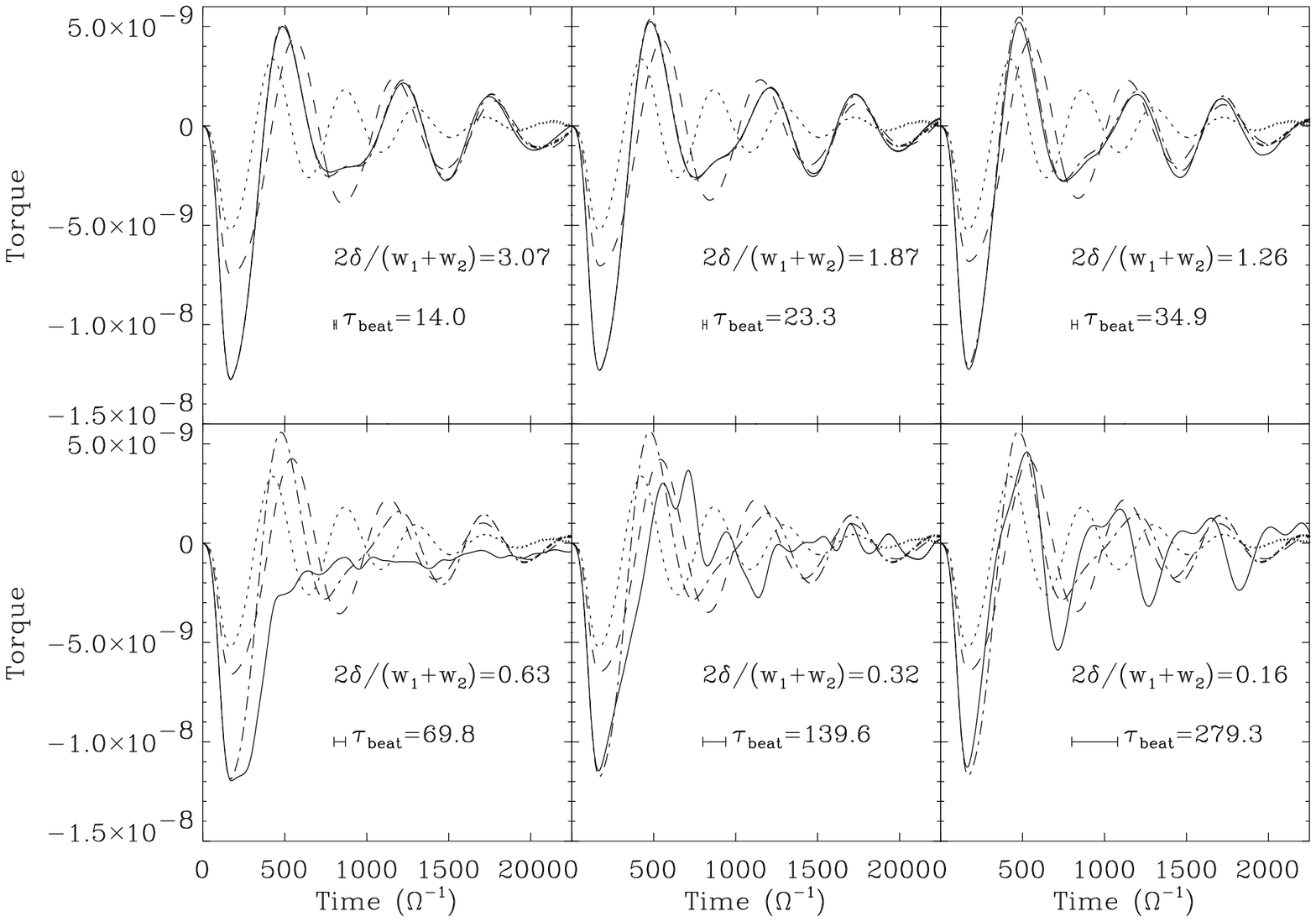}
\caption{\label{fig:seven}%
Corotation torque of a pair of corotation resonances in the inviscid
case, as a function of time. The curve styles are the same as in the
previous figure.}
\end{figure}

In Figure~\ref{fig:seven} we show results for six values of the
separation ($s=500,300,200,100,50,25$) of the inviscid case. Again the
solid and dashed-dotted curves are almost indistinguishable by eye
when the resonances do not overlap ($\overline\delta >1$,
corresponding to the top panels), while they are markedly different
when they overlap ($\overline\delta < 1$, bottom panels).  All curves
slowly decay in absolute value.  This is normal for the one-resonance
torques and their sum, as the corotation torque in that case tends to
saturate, and it is reasonable to expect it also to saturate, on the
average, in the two-resonance case. A quasi-periodic torque modulation
is apparent for the closest case (right-bottom plot), with a
pseudo-period that corresponds to $2\tau_{\rm beat}$.

\placefigure{fig:seven}

\subsection{Systematic exploration}
\begin{figure}
\plotone{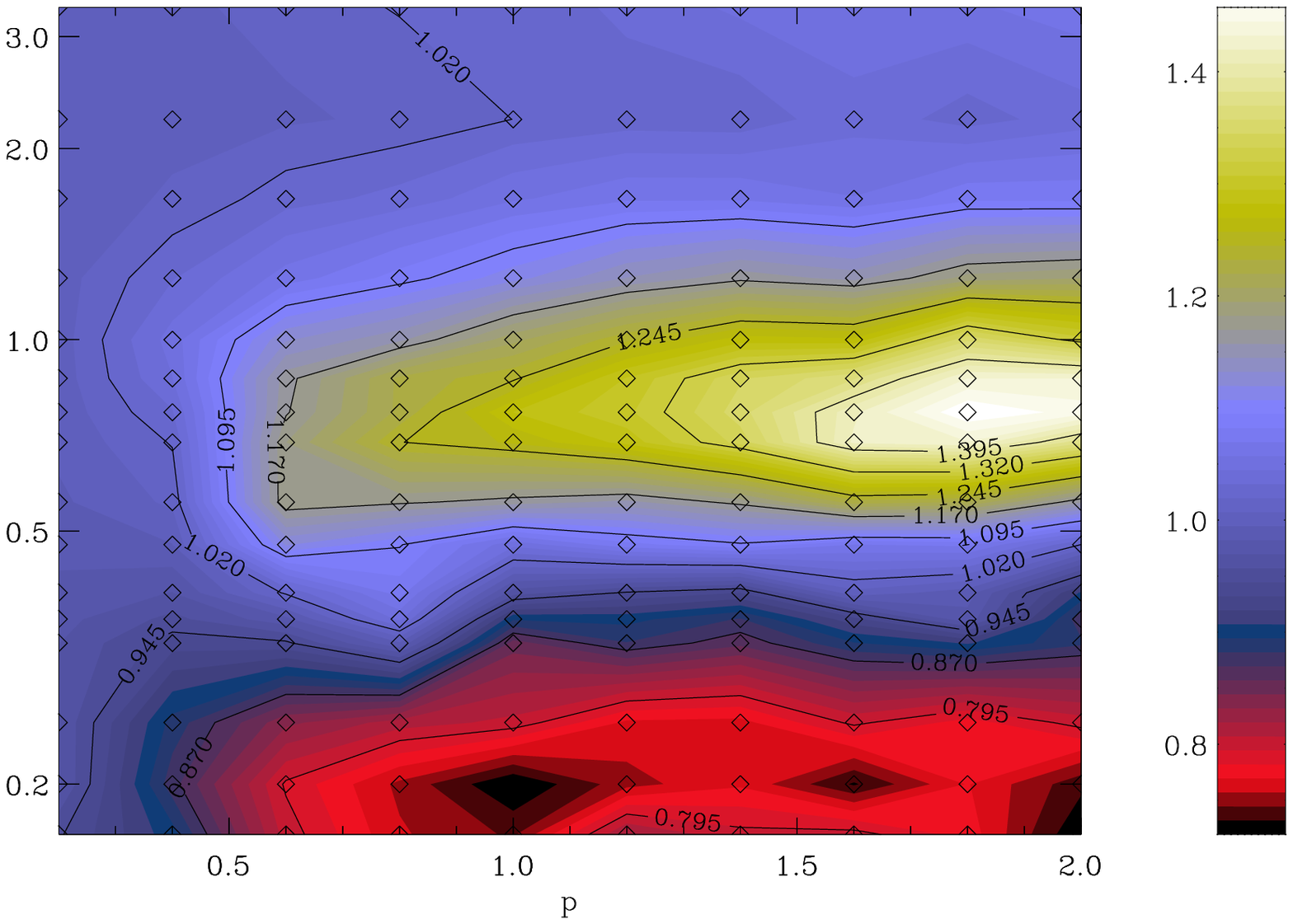}
\caption{\label{fig:eight}%
Ratio of the two-resonance torque to the sum of the torque of the
corresponding resonances considered separately, as a function of the
saturation parameter $p$ and the overlap parameter. Viscosity
increases to the left, and the separation increases to the top.}
\end{figure}

We have performed calculations similar to the illustrative cases of
two corotation resonances of the previous paragraphs, with a more
systematic exploration of the (separation, saturation) parameter
space.  Our aim is to check the following trends:
\begin{enumerate} 
\item 
The   two-resonance torque value   is  the  sum  of the  one-resonance
torques, whenever  the viscosity  is high  or   the resonances do  not
overlap.
\item 
The two-resonance torque value of overlapping resonances saturates at
low viscosities, in a fashion that is similar to the saturation of a
one-resonance torque.
\item 
The time-averaged torque value of two overlapping resonances for a
small or moderate viscosity may sensibly differ from the sum of the
isolated resonances torques, but it never exceeds the sum of the fully
unsaturated isolated resonance torques.
\end{enumerate}
We therefore set up the following series of calculations:
\begin{itemize}
\item 
We consider two fixed corotation resonances, with frequencies and
azimuthal wavenumbers respectively $\Omega=1.015$ and $m=3$, and
$\Omega'=0.985$ and $m'=2$.  These resonances have the same forcing
potential amplitude $\psi$, which we vary. The disk still has
$H/r=0.01$, the mesh has $N_r=240$ by $N_\phi=90$ zones, the inner
mesh boundary is at $R_{\rm min}=0.84$, the outer mesh boundary is at
$R_{\rm max}=1.16$. The potential is turned on on the time-scale
$\tau=20\;\Omega^{-1}$. We use damping boundary conditions.
\item 
The amplitude of the forcing potential is chosen such that the
parameter ${\overline\delta}^{-1}=(w_1+w_2)/2\delta$ takes the
following values: $0.3$, $0.45$, $0.6$, $0.8$, $1.0$, $1.15$, $1.3$,
$1.45$, $1.8$, $2.1$, $2.5$, $2.75$, $3.0$, $4.0$, $5.0$, $6.0$, i.e.
from well separated resonances to strongly overlapping resonances. In
the above expression $w_1$ and $w_2$ are evaluated as if the resonance
frequencies were both $\Omega=1$, therefore $w_1=w_2$ since the
libration island width, in the ballistic approximation, does not
depend on $m$.
\item 
The viscosity is varied such that the saturation parameter, given by
equation~(\ref{eq:sat}), of the ``mean'' fictitious resonance that
would have $\Omega=1$, the same forcing potential as the two resonances
considered, and $m=2.5$, would take the values $p=0.2, 0.4, 0.6,
\ldots 2.0$.
\item 
Each calculation is performed over a time that is larger than the
libration time of each resonance, the beating time between them, and
the viscous time across each libration zone. The torque is then
measured and averaged over the last beating time.
\item 
The same procedure is repeated for two other calculations in which
only one of the resonances is present, and which is performed over
exactly the same time, and again the torque is measured and
time-averaged over the same temporal window (although the beating time
is then meaningless).  The whole calculation therefore amounts in
total to $3\times 16\times 10=480$ such runs.
\end{itemize}
In Figure~\ref{fig:eight} we plot the ratio of the time-averaged
two-resonance torque to the sum the one-resonance torques, as a
function of viscosity and resonance separation.  As expected, the
result is very close to unity near the top and left axes, which
respectively correspond to the well separated case and the highly
viscous case. This confirms the expectation~1 above.

\placefigure{fig:eight}

The fact that the ratio plotted in Figure~\ref{fig:eight} does not
differ from unity by more than roughly $40$\% confirms the
expectation~2 above, i.e. the saturation curve of total torque
vs. $p$, taken at any given value for the overlap parameter, does not
significantly differ from the sum of the saturation curves of the
corresponding resonances, considered as isolated.

\begin{figure}
\plotone{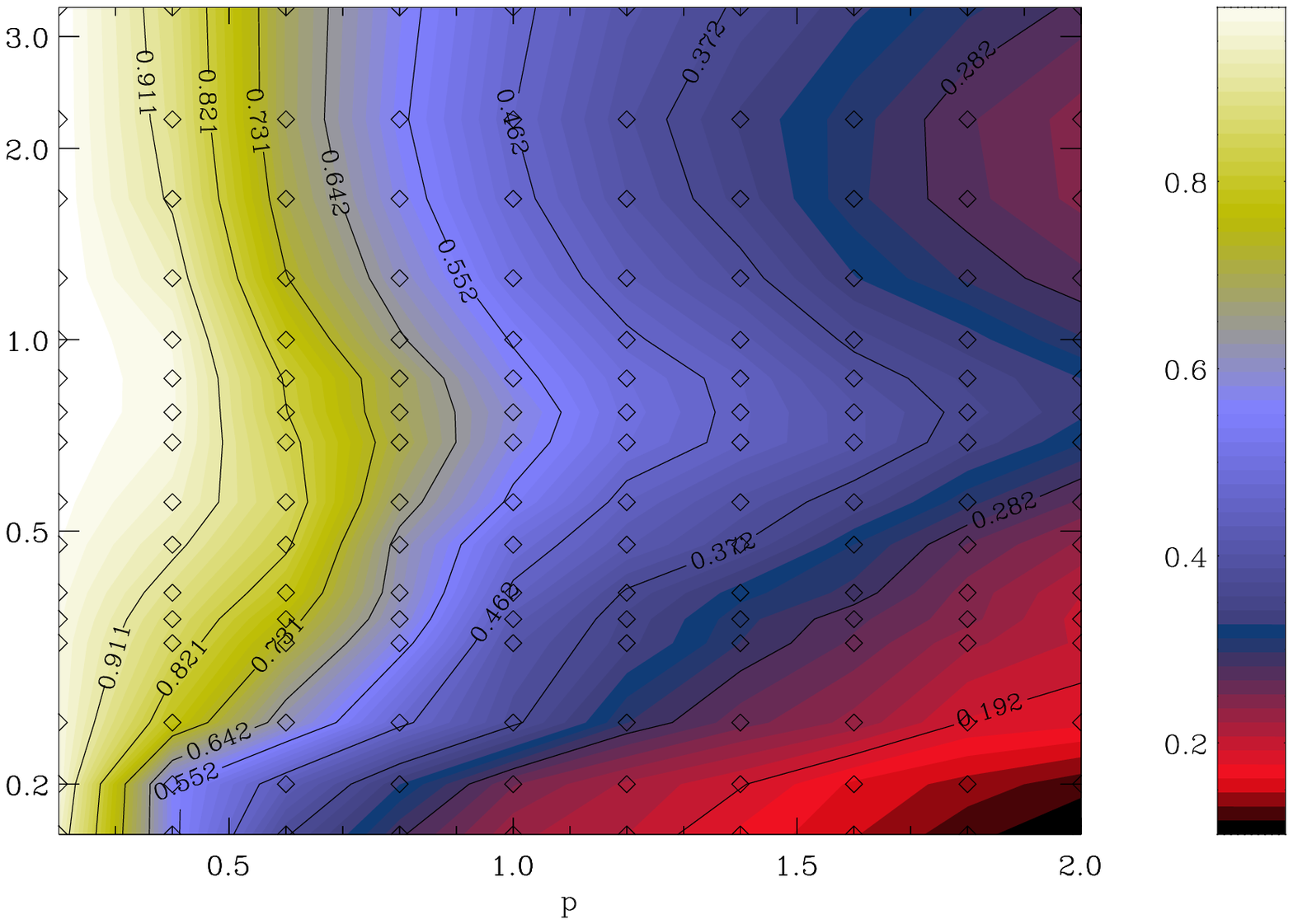}
\caption{\label{fig:nine}%
Ratio of the two-resonance torque to the sum of the torques of the
corresponding fully unsaturated resonances considered separately, as a
function of the saturation parameter $p$ and the overlap parameter.
Viscosity increases to the left, and the separation increases to the
top.}
\end{figure}

\placefigure{fig:nine}

We plot in Figure~\ref{fig:nine} the ratio of the two-resonance torque
value to the sum of the fully unsaturated torque values of both
resonances, as a function of viscosity and resonance separation. This
ratio can be seen to never exceed unity, which confirms expectation~3
above.

\section{Saturation of a corotation resonance overlapping a Lindblad resonance}

As the first-order corotation resonances lie on top of principal
Lindblad resonances, we have addressed the restricted problem of the
saturation of one corotation resonance that exactly overlaps a
Lindblad resonance.  We have examined the consequences of this overlap
in two different situations:
\begin{itemize}
\item 
The Lindblad torque is small enough so that it does not significantly
affect the azimuthally averaged radial profile of surface
density. This allows us to compare directly the corotation torque
value with its value in the case no Lindblad resonance is present.
\item 
The Lindblad torque is large and significantly redistribute the disk
material and thus it considerably alters the slope of vortensity
across the resonance. In that case, a direct comparison with a single
corotation calculation is not possible. Rather, we measure the slope
of vortensity across the resonance and compare the value of the
unsaturated resonance to the Goldreich \& Tremaine estimate.
\end{itemize}

\subsection{Small Lindblad torque}

\begin{figure}
\plotone{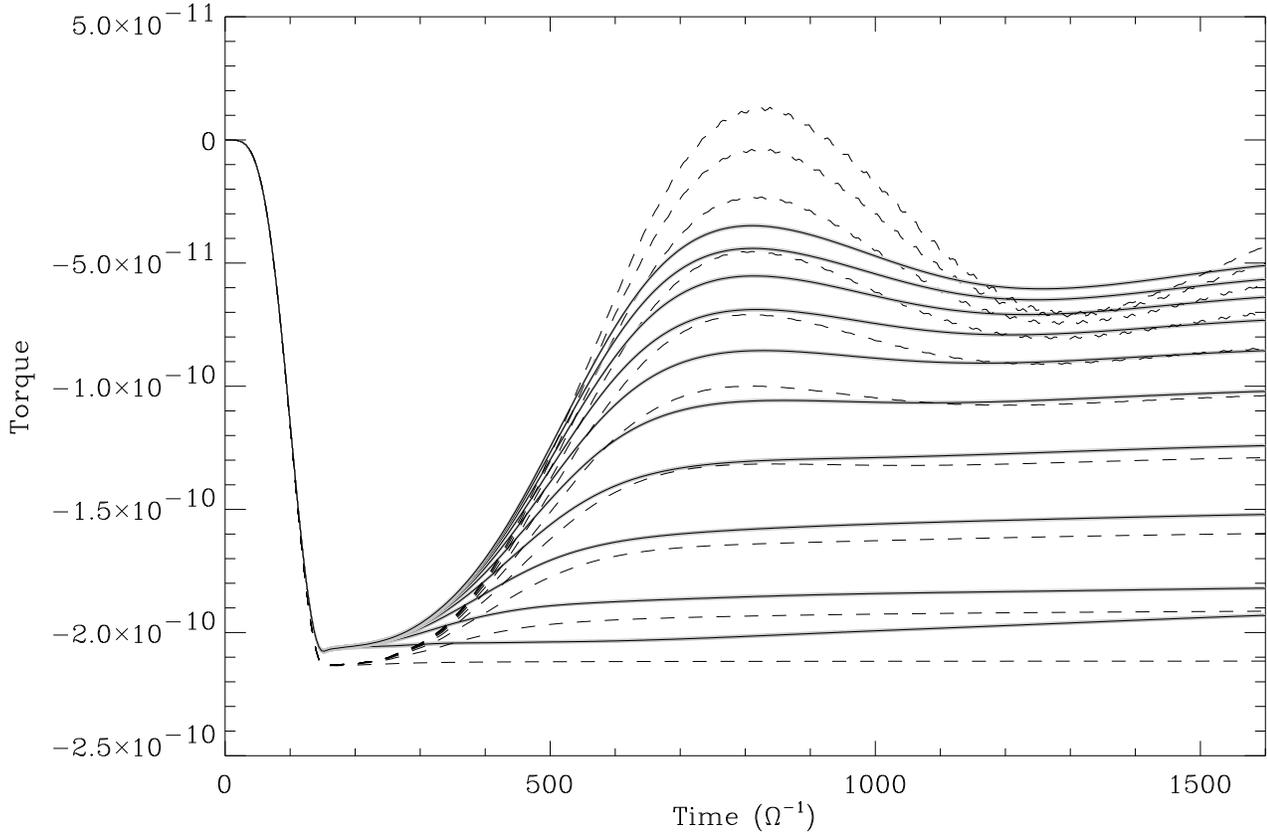}
\caption{\label{fig:ten}%
Corotation torque of the resonance described in text, as a function a
time, for different values of the viscosity (low viscosities are at
the top), for the case of an overlapping Lindblad resonance (solid
curves) and for the case of an isolated resonance (dashed curves).}
\end{figure}

\placefigure{fig:ten}

\begin{figure}
\plotone{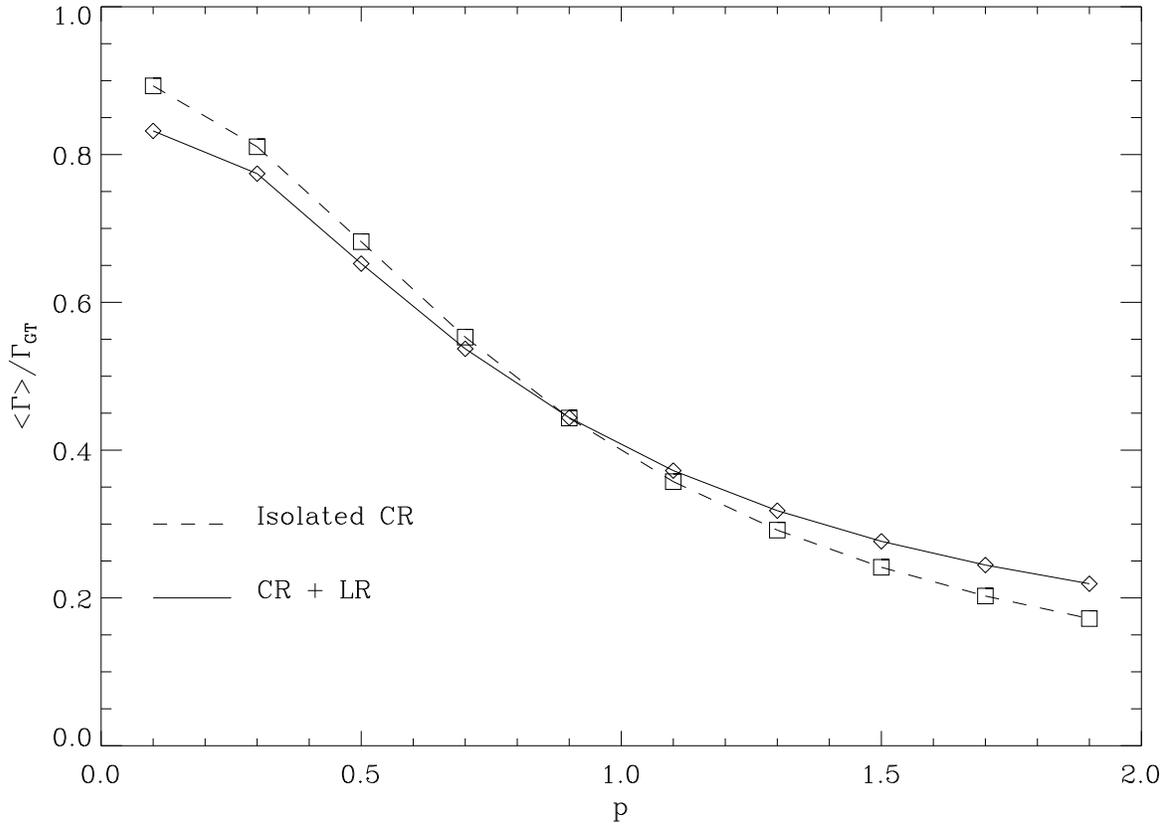}
\caption{\label{fig:eleven}%
Dimensionless corotation torque as a function of the saturation
parameter $p$, for the case of a mixed corotation/Lindblad resonance
(solid line) and for the case of an isolated corotation resonance
(dashed line). }
\end{figure}

\placefigure{fig:eleven}

We consider a corotation resonance excited by a potential with
wavenumber $m=3$, frequency $\Omega = 1$ and amplitude $\psi=2\cdot
10^{-6}$, which has exactly the same location as a Lindblad resonance
excited by a potential with wavenumber $m'=2$, frequency $\Omega'=3/2$
and amplitude $\psi'=5\cdot 10^{-5}$.  For the case of an eccentric
planet, this corresponds to a slow first-order corotation resonance
overlapping a principal outer Lindblad resonance.  The reason why we
consider such a low-amplitude forcing potential for the corotation
resonance is that we wish the onset of possible non-linear effects
between the corotation and Lindblad resonances to be as low as
possible.

Our mesh extends from $R_{\rm min}=0.84$ to $R_{\rm max}=1.16$, its
resolution is $N_r=1000$ by $N_\phi=90$. The radial resolution is
large in order to have a satisfactory resolution of the narrow
libration islands of the corotation resonance. The potentials are
turned on over the time-scale $\tau=150\;\Omega^{-1}$.

In Figure~\ref{fig:ten} we plot the torque value of the corotation
resonance as a function of time for different values of the viscosity,
corresponding to an arithmetic sequence of the saturation parameter
$p$, for a mixed corotation/Lindblad resonance and for the case of an
isolated corotation resonance with the same characteristics.

Both series of curves exhibit very similar trends.  The torque of the
isolated resonance tends to oscillate more than the one overlapping a
Lindblad resonance, at low viscosity. One can also notice the very
long libration time, due to the narrowness of the libration islands.

\begin{figure}
\plottwo{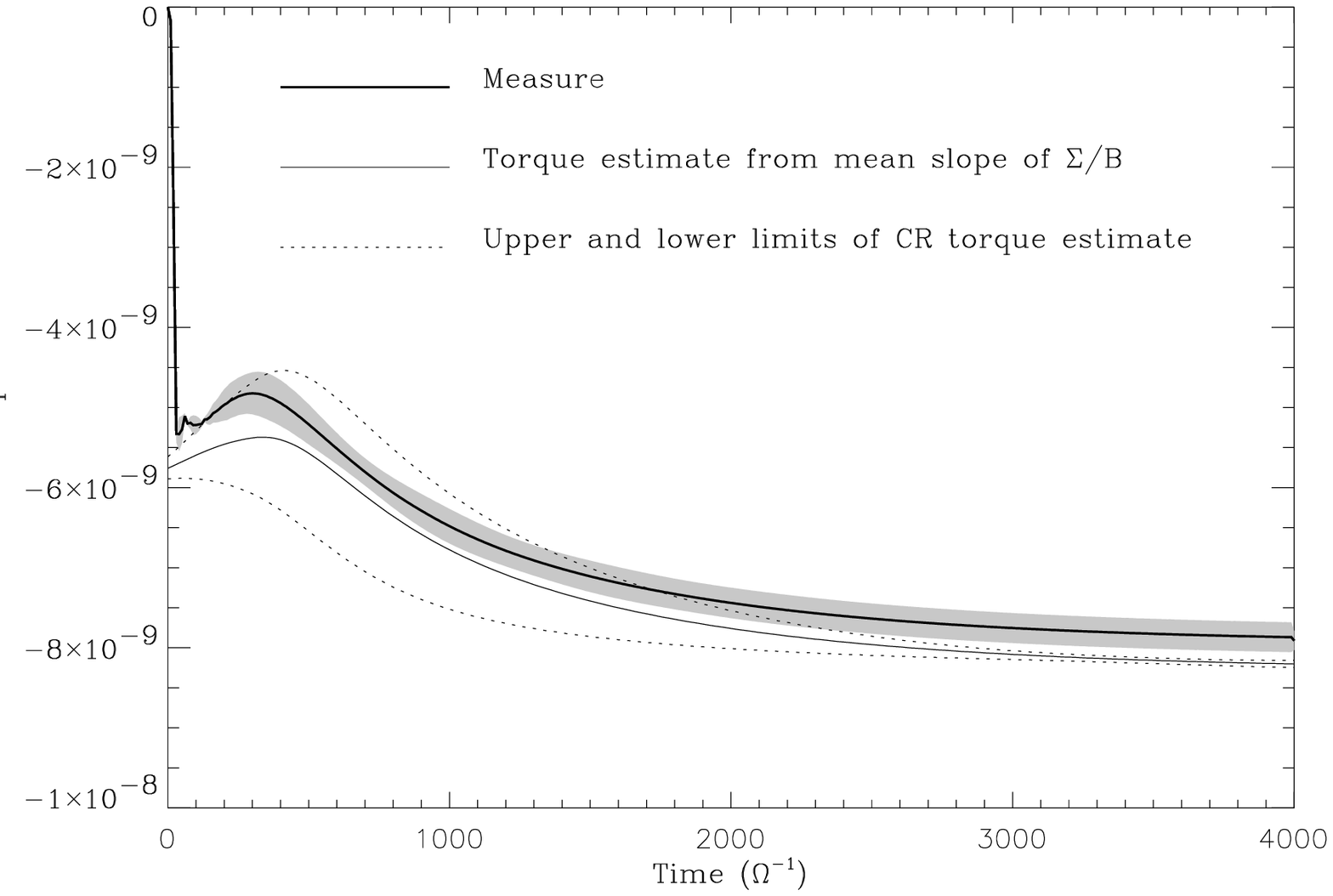}{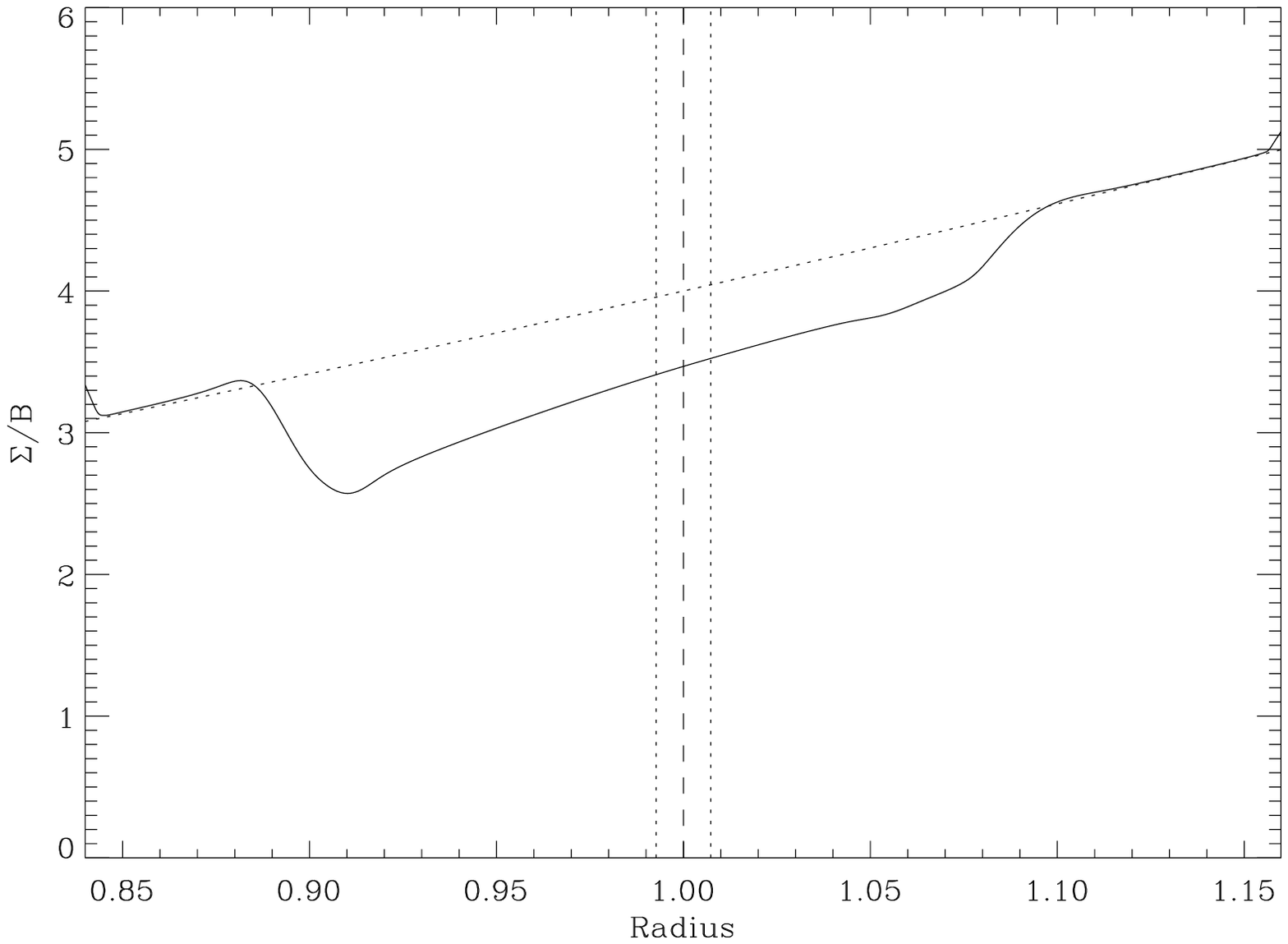}
\caption{\label{fig:twelve}%
Torque of a corotation resonance overlapping a Lindblad resonance
(thick solid line).  The Goldreich \& Tremaine estimate of that torque
using the mean value of the slope of $\Sigma/B$ across the corotation
resonance width is shown with the thin solid line, while the extreme
values of this estimate, which correspond to the extrema of the slope
of $\Sigma/B$ across the resonance width, are shown as dotted lines.
The shaded area corresponds to the excursion of the torque value,
which is modulated on the short time-scale $\tau_{\rm beat}$. The
right panel shows the vortensity profile at $t=0$ (tilted dotted line)
and at $t=2000\;\Omega^{-1}$. The common position of the resonances is
marked with the dashed line.  The equidistant vertical dotted lines
show the limit of the corotation libration islands -- determined as if
it was isolated.}
\end{figure}

\placefigure{fig:twelve}

In Figure~\ref{fig:eleven} we plot the dimensionless torque as a
function of the saturation parameter, obtained from the time average
of the torque values presented in Figure~\ref{fig:ten}, over the
interval $800\;\Omega^{-1} < t < 1600\;\Omega^{-1}$.

The time-averaged torque estimate at low viscosity, especially for the
isolated resonance, may be slightly biased by the fact that it has not
enough time to reach a steady saturation level over the full duration
of the calculation. The impact of this bias is however small, and the
conclusion that can be drawn from the examination of
Figure~\ref{fig:eleven} is that the torque saturation properties are
hardly affected by the presence of the underlying Lindblad resonance.
The tiny difference between both curves can be attributed either to
the relatively small number of libration times over which the time
average is performed (at low viscosity) or to the slight radial
redistribution of the disk material under the action of the Lindblad
torque.  The tiny change of the vortensity slope, defined as
$-d\log(\Sigma/B)/d\log r$ (where $B$ is the second Oort's constant),
can indeed be estimated to be of the order of a few percent,
compatible with the discrepancy of the two saturation curves.

\subsection{Large Lindblad torque}

We have also performed a test calculation in which the forcing
potential amplitude of the Lindblad resonance is much larger, so that
it modifies the surface density profile, in a reduced version of what
occurs at a gap edge. We cannot vary the viscosity in this kind of
calculation, as we want the surface density profile to converge
toward the redistributed profile in a reasonable amount of time.  At
viscosities low enough to significantly saturate the corotation
torque, the viscous time-scale across the excavated region becomes
prohibitively large.  We therefore just compare the corotation torque
value to the Goldreich \& Tremaine estimate, obtained by measuring the
slope of $\Sigma/B$ across the resonance width at any instant in
time. In Figure~\ref{fig:twelve}, the left panel shows the comparison
between this estimate and the actual corotation torque value.  They
agree to within a few percent, and the actual torque value is slightly
smaller than the Goldreich--Tremaine estimate.  The parameters that we
adopted for this calculation are:
\begin{itemize}
\item 
$\psi=10^{-5}$, $m=3$ and $\Omega=1$ for the corotation resonance,
\item 
$\psi'=10^{-4}$, $m'=2$ and $\Omega'=1/2$ for the Lindblad resonance.
This frequency implies that it is an ILR.  If we wanted to strictly
reproduce the overlap between a first-order corotation resonance and
a principal inner Lindblad resonance, we should have $m_{\rm
ILR}=m_{\rm CR}+1$. We have chosen instead $m_{\rm ILR}=m_{\rm CR}-1$,
which allows us to better resolve the wave launched at the Lindblad
resonance for a given value of $N_\phi$.  We anticipate that our
conclusions are not affected by this choice.
\end{itemize}

\section{Low-resolution issues}
The above sections essentially validate the previous work of Ogilvie
\& Lubow (2003) on the saturation of an isolated corotation
resonance, in the context of the first-order corotation resonances of
an eccentric giant planet, which are not isolated. The numerical
simulations performed thus far on the problem of the eccentricity
excitation of a giant planet did not exhibit any eccentricity growth
for Jupiter-sized giant planets (Papaloizou et al. 2001). While there
may be a number of reasons for that, we show in this section how the
corotation torque estimate is mangled at very low resolution,
i.e. when the width of the libration islands is of the order of the
mesh radial size.  The width of a first-order corotation resonance is
given by:
\begin{equation}
  \frac{W}{r} \approx 4.1 (C_m^\pm m e q)^{1/2},
\end{equation}
where $q$ is the mass ratio and $C_m^\pm$ is a factor of order unity
given in Tables 1 and 2 of Ogilvie \& Lubow (2003). Assuming one has a
Jupiter-sized planet orbiting a solar mass star ($q=10^{-3}$) with
eccentricity $e=0.01$, the resonance width is therefore
$W/r\approx0.013(C_m^\pm m)^{1/2}$. As a point of comparison, in run
N1 in Papaloizou et al. (2001), which corresponds to a Jupiter-sized
giant planet, the radial resolution is $\Delta r/r=0.043$.

\begin{figure}
\plotone{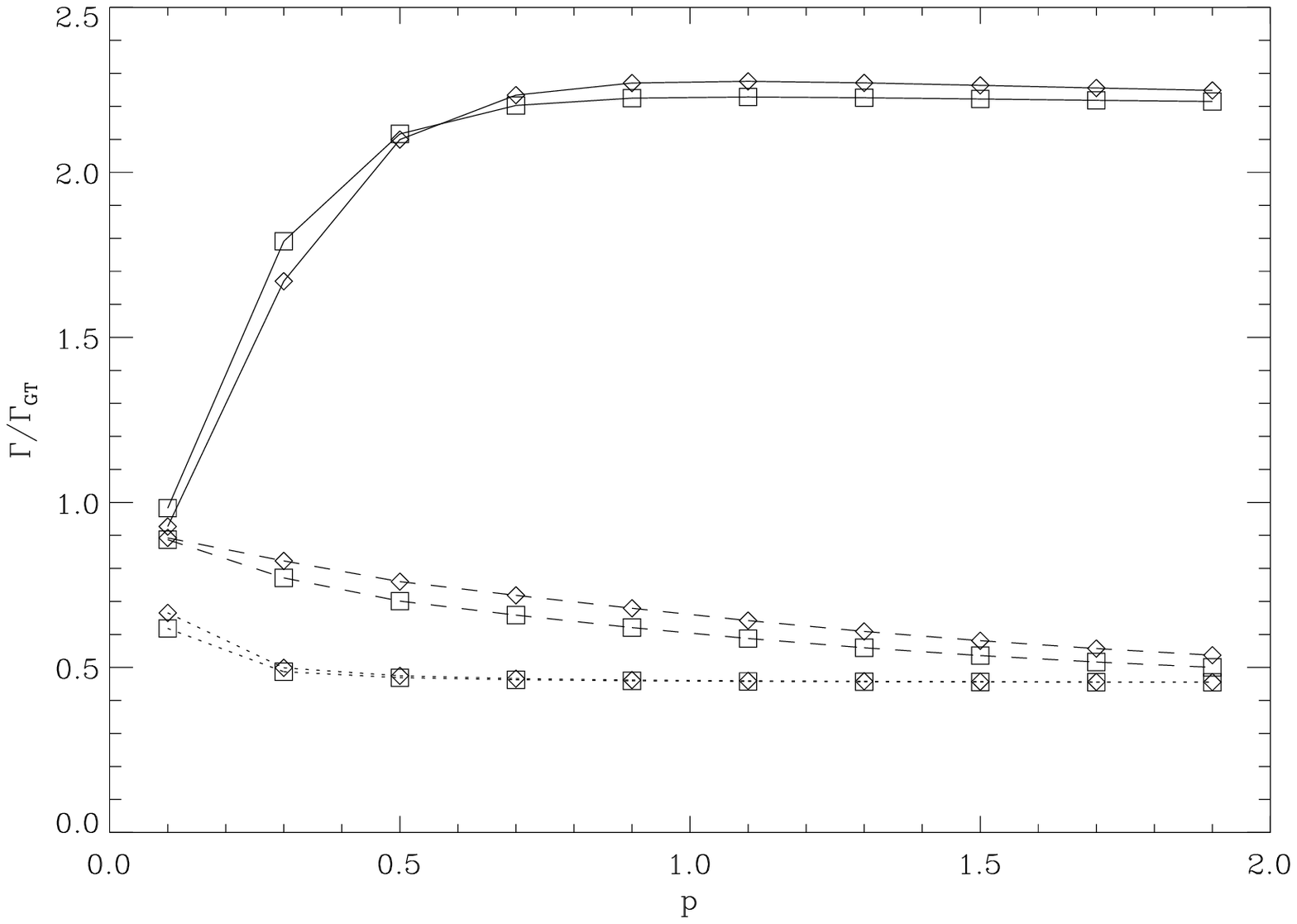}
\caption{\label{fig:thirteen}
Dimensionless torque of an isolated corotation resonance as a
function of viscosity, for three different resolutions.  }
\end{figure}

\placefigure{fig:thirteen}

Figure~\ref{fig:thirteen} shows the results (as before, diamonds are
for a uniform viscosity prescription while squares are for a
$\nu\propto\Sigma^{-2/3}$ viscosity prescription). The solid lines are
for a resolution $W/\Delta r = 1.0$, the dotted one $W/\Delta r = 0.5$
and the dashed one $W/\Delta r = 2.0$. This shows again that the value
of the fully unsaturated corotation torque is relatively well
reproduced even at very low resolution, but the saturated resonant
torque is completely mangled at low resolution.  The case $W=\Delta r$
is particularly striking, since in this case one gets a torque value
larger than the Goldreich \& Tremaine estimate.  Therefore in a
low-resolution simulation of an embedded giant planet, as soon as the
eccentricity gets to values which give the libration islands a size
comparable to the radial resolution, these effects could lead to a
strong spurious damping of the eccentricity, which would therefore be
bounded by small values ($O[0.01]$, depending on the resolution).

\section{Conclusions}

By means of reduced two-dimensional numerical simulations, involving
two resonances at the same time and each resonance separately in
order to compare the isolated case to the two-resonance case, we reach
the following conclusions:

\begin{itemize}
\item
The corotation torque of two overlapping resonances coincides with the
sum of the corotation torques of each resonance, considered as
isolated, either when the resonances do not overlap (their separation
is larger than the sum of the half-widths of their libration islands)
or when the viscosity is large (fully unsaturated resonances).
\item
When the resonances overlap and are partially saturated, the
two-resonance torque may differ at most by a few tens of percent from
the sum of the isolated torques, evaluated for the same viscosity.
This implies that a given degree of saturation of a non-isolated
resonance is reached for a value of the viscosity of the same order of
magnitude as the viscosity needed to achieve the same degree of
saturation for the isolated resonance.
\item
The time-averaged two-resonance torque never exceeds the sum of the
Goldreich \& Tremaine estimates of each resonance.
\item
In the case where a corotation resonance overlaps a weak Lindblad
resonance (that does not significantly redistribute the vortensity
profile), the time-averaged torque value differs from the isolated
case by at most a few percent, whatever the value of the viscosity.
\item
In the case where a corotation resonance overlaps a strong Lindblad
resonance, for which our calculations are limited to high-viscosity
situations, the time-averaged torque value corresponds to the
Goldreich \& Tremaine estimate, i.e. to the torque of the unsaturated
corotation resonance, considered as isolated.
\item 
The only case in which we have observed a corotation torque value that
is larger than the Goldreich \& Tremaine estimate (which therefore
jeopardizes an eccentricity excitation for the set of parameters
quoted in the introduction) is when the mesh resolution is too low.
This is therefore a numerical artifact, and we conclude that a high
resolution is needed in the region of first-order corotation resonance
(i.e. close to the gap edges) in order to observe an eccentricity
excitation, with libration islands spanning at least about $10$ zones
radially.
\end{itemize}

Finally, we emphasize that the saturation of the first-order
corotation resonances is only one of the factors determining whether
the eccentricity of a giant protoplanet grows or decays through its
interaction with the disk.  An aspect of this problem that tends to be
overlooked is the secular exchange of eccentricity between the planet
and the disk, which may occur on a time-scale of thousands of orbital
periods. The mechanism responsible for this exchange is sometimes
referred to as the apsidal resonance (Goldreich \& Sari 2003),
although this resonance is so broad in a nearly Keplerian disk that we
prefer to avoid this denomination.  When a planet is placed on an
elliptical orbit in a circular disk, its eccentricity should initially
decrease as eccentricity is imparted to the disk.  This exchange can,
in principle, occur in a way that conserves the angular momentum
deficit of the coupled disk--planet system, in which case the
eccentricity of the planet will be restored later in the cycle of
secular exchange.  Therefore, a transient decrease of the planet's
eccentricity in such circumstances should not be confused with
eccentricity damping.  In the analogous problem of the secular
inclination dynamics of coupled disk--planet systems, Lubow \& Ogilvie
(2001) have defined a positive-definite measure of the bending
disturbance of the system, equivalent to an angular momentum deficit,
which is conserved in the secular exchange and evolves only through
the effects of mean-motion resonances and viscous damping.  Papaloizou
(2002) has also treated aspects of the eccentricity dynamics of
coupled disk--planet systems, while Ogilvie (2001) has drawn attention
to the subtle issue of whether eccentricity is damped or excited by
viscous or turbulent stresses in the disk, and at what rate.

A fully self-consistent calculation therefore involves many additional
complications, such as the treatment of the force arising from the
material inside of the Roche lobe, the role of self-gravity, and the
eccentric coupling between the disk and the planet through the apsidal
waves. It will be presented in a forthcoming work.

{}

\clearpage

\begin{deluxetable}{lcc}
\tablecaption{\label{tab:reslist}%
List of the locations of resonances and their action on the
eccentricity. Bracketed terms correspond to co-orbital resonances
($\Omega=\Omega_p$) which play no role when the planet clears a gap.
}
\tablehead{
\colhead{Resonance} &\colhead{$(\frac r{r_p})^{-2/3}=\frac\Omega{\Omega_p}$} & \colhead{$(\frac{1}{e})(\frac{de}{dt})_{\rm res}$}}
\startdata
Principal OLR & $\frac m{m+1}$ 	&	$+$ \\
Principal ILR & $\frac m{m-1}$ 	&	$-$ \\
Principal CR &  $1$ 		&	$[?]$ \\
\hline
Fast first-order OLR & 1 	&	$[-]$ \\
Fast first-order ILR & $\frac {m+1}{m-1}$ &	$+$ \\
Fast first-order CR  & $\frac{m+1}{m}$ 	&	$-$ \\
\hline
Slow first-order OLR & $\frac{m-1}{m+1}$ &	$+$ \\
Slow first-order ILR & $1$ 		&	$[-]$ \\
Slow first-order CR  & $\frac{m-1}{m}$  &	$-$ \\
\enddata
\end{deluxetable}

\clearpage

\clearpage

\clearpage

\clearpage

\clearpage

\clearpage

\clearpage

\clearpage

\clearpage

\clearpage

\clearpage

\clearpage

\begin{thebibliography}{}
\bibitem[2003]{akh03} 
d'Angelo, G., Kley, W., \& Henning, T.\ 2003, ApJ, 586, 540 
\bibitem[2003]{b03} 
Bate, M.\ R., Lubow, S.\ H., Ogilvie, G.\ I., \& Miller, K.\ A.\ 2003, MNRAS, 341, 213 
\bibitem[2001]{f01}
Ford, E.\ B., Havlickova, M., \& Rasio, F.\ A.\ 2001, Icarus, 150, 303
\bibitem[2003]{fg03} 
Goldreich, P.\ \& Sari, R.\ 2003, ApJ, 585, 1024 
\bibitem[1979]{gt79} 
Goldreich, P.\ \& Tremaine, S.\ 1979, ApJ, 233, 857 
\bibitem[1980]{gt80}
Goldreich, P.\ \& Tremaine, S.\ 1980, ApJ, 241, 425 
\bibitem[1997]{h97}
Holman, M., Touma, J., \& Tremaine, S.\ 1997, Nature, 386, 254 
\bibitem[1998]{k98}
Kley, W.\ 1998, A\&A, 338, 37
\bibitem[2001]{lo01}
Lubow, S.\ H.\ \& Ogilvie, G.\ I.\ 2001, ApJ, 560, 997
\bibitem[2002]{m02}
Masset, F.\ S.\ 2002, A\&A, 387, 605
\bibitem[2001]{m01}
Masset, F.\ S.\ 2001, ApJ, 558, 453
\bibitem[2000]{m00}
Masset, F.\ S.\ 2000a, A\&AS, 141, 165
\bibitem[1997]{m97}
Mazeh, T., Krymolowski, Y., \& Rosenfeld, G.\ 1997, ApJ, 477, L103
\bibitem[1987]{mvs87}
Meyer-Vernet, N. \& Sicardy, B.\ 1987, Icarus, 69, 157
\bibitem[2000]{n00}
Nelson, R.\ P., Papaloizou, J.\ C.\ B., Masset, F.\ S., \& Kley, W.\ 2000,
MNRAS, 318, 18
\bibitem[2001]{o01}
Ogilvie, G.\ I.\ 2001, MNRAS, 325, 231
\bibitem[2003]{ol03} 
Ogilvie, G.\ I.\ \& Lubow, S.\ H.\ 2003, ApJ, 587, 398
\bibitem[2002]{pap02}
Papaloizou, J.\ C.\ B.\ 2002, A\&A, 388, 615
\bibitem[2001]{pap01}
Papaloizou, J.\ C.\ B., Nelson, R.\ P., \& Masset, F.\ S. 2001, A\&A, 366, 263
\bibitem[1996]{rf96}
Rasio, F.\ A.\ \& Ford, E.\ B.\ 1996, Science, 274, 954
\bibitem[1992]{sn92} 
Stone, J.\ M.\ \& Norman, M.\ L.\ 1992, ApJS, 80, 753 
\bibitem[1977]{vl77} 
van Leer, B.\ 1977, Journal of Computational Physics, 23, 276 
\end{thebibliography}
\end{document}